\documentclass[10pt]{article}
\usepackage[T2A,T1]{fontenc}
\usepackage[latin9]{inputenc}
\usepackage{amsmath}
\usepackage{amssymb}
\usepackage{graphicx}
\usepackage{wasysym}
\usepackage[unicode=true,
 bookmarks=true,bookmarksnumbered=true,bookmarksopen=true,bookmarksopenlevel=3,
 breaklinks=false,pdfborder={0 0 0},pdfborderstyle={},backref=false,colorlinks=false]
 {hyperref}
\hypersetup{pdftitle={A scaling law from discrete to continuous solutions of channel capacityproblems in the low-noise limit},
 pdfauthor={Michael C. Abbott & Benjamin B. Machta},
 pdfsubject={An analog communication channel typically achieves its full capacitywhen the distribution of inputs is discrete, composed of just $K$symbols, such as voltage levels or wavelengths. As the effective noiselevel goes to zero, for example by sending the same message multipletimes, it is know that optimal codes become continuous. Here we derivea scaling law for the optimal number of symbols in this limit, findinga novel rational scaling exponent. The number of symbols in the optimalcode grows as $\log K\sim I^{4/3}$, where the channel capacity $I$increases with decreasing noise. The same scaling applies to otherproblems equivalent to maximizing channel capacity over a continuousdistribution.},
 pdfborderstyle={},citecolor={darkblue},urlcolor={darkblue}}

\makeatletter
\pdfoutput=1

\usepackage{xcolor}
\usepackage{pdfcolmk}
\usepackage{units}
\usepackage{wasysym}
\PassOptionsToPackage{normalem}{ulem}
\usepackage{ulem}

\providecolor{lyxadded}{rgb}{0,0,1}
\providecolor{lyxdeleted}{rgb}{1,0,0}

\usepackage{M12}

\renewcommand{\vec}[1]{\oldvec{#1}}

\usepackage{pgfplots}

\def\easycyrsymbol#1{\mathord{\mathchoice
  {\mbox{\fontsize\tf@size\z@\usefont{T2A}{\rmdefault}{m}{n}#1}}
  {\mbox{\fontsize\tf@size\z@\usefont{T2A}{\rmdefault}{m}{n}#1}}
  {\mbox{\fontsize\sf@size\z@\usefont{T2A}{\rmdefault}{m}{n}#1}}
  {\mbox{\fontsize\ssf@size\z@\usefont{T2A}{\rmdefault}{m}{n}#1}}
}}

\newcommand{\popt}{p_\star}
\newcommand{\pjeff}{p_\mathrm{J}}

\let\originalleft\left
\let\originalright\right
\renewcommand{\left}{\mathopen{}\mathclose\bgroup\originalleft}
\renewcommand{\right}{\aftergroup\egroup\originalright}

\AtBeginDocument{
  
}

\makeatother

\begin{document}
\newcommand{\markoneempty}{\textcolor{darkblue}{$\,\lozenge$}}
\newcommand{\markone}{\textcolor{darkblue}{$\,\blacklozenge$}}
\newcommand{\marktwoempty}{\textcolor{darkred}{\Large $\circ$}}
\newcommand{\marktwo}{\textcolor{darkred}{\Large $\bullet$}}
\newcommand{\markthree}{\textcolor{darkgreen}{$\blacksquare$}}

\title{A Scaling Law From Discrete to Continuous Solutions of Channel Capacity
Problems in the Low-Noise Limit}
\author{Michael C. Abbott\alabel{\markone} \emph{\&} \  Benjamin B. Machta\alabel{\marktwo}
\address[\markoneempty]{Institute of Physics, Jagiellonian University,
\\
Ulica \L ojasiewicza 11, 30-348 Kraków, Poland.}\address[\markone]{Holographic
QFT Group, MTA Wigner Research Centre for Physics, \\
Konkoly-Thege Mikl\'{o}s u. 29-33, 1121 Budapest, Hungary.}\address[\marktwoempty]{Lewis-Sigler
Institute and Department of Physics, \\
Princeton University, Princeton, NJ 08544, USA.}\address[\marktwo]{Department
of Physics and Systems Biology Institute, \\
Yale University, New Haven, CT 06520, USA.}\address{michael.abbott@wigner.mta.hu,
benjamin.machta@yale.edu}}
\date{v1: October 2017\\
v2: \href{https://doi.org/10.1007/s10955-019-02296-2}{J. Stat. Phys.}
April 2019}
\maketitle
\begin{abstract}
An analog communication channel typically achieves its full capacity
when the distribution of inputs is discrete, composed of just $K$
symbols, such as voltage levels or wavelengths. As the effective noise
level goes to zero, for example by sending the same message multiple
times, it is known that optimal codes become continuous. Here we derive
a scaling law for the optimal number of symbols in this limit, finding
a novel rational scaling exponent. The number of symbols in the optimal
code grows as $\log K\sim I^{4/3}$, where the channel capacity $I$
increases with decreasing noise. The same scaling applies to other
problems equivalent to maximizing channel capacity over a continuous
distribution.
\end{abstract}

\section{Introduction\label{sec:Introduction}}

Information theory is concerned with communication in the presence
of noise \cite{Shannon:1948wk}. A noisy channel may be described
by the probability distribution $p(x|\theta)$ over received messages
$x\in X$, for a given signal $\theta\in\Theta$. The mutual information
between input and output $I(X;\Theta)$ depends not only on the channel,
but also on the distribution of input signals $p(\theta)$. A choice
of communication code implies a choice of this input distribution,
and we are interested in $\popt(\theta)$ which maximizes $I$, which
is then precisely the channel capacity.

Some channels are used repeatedly to send independent signals, as
for example in telecommunications. One surprising feature of the optimal
distribution $\popt(\theta)$ in this context is that it is usually
discrete: Even when $\theta$ may take on a continuum of values, the
optimal code uses only finite number $K$ of discrete symbols. In
a sense the best code is digital, even though the channel is analog.

The opposite limit of the same problem has also been studied, where
effectively a single signal $\theta$ is sent a large number number
of times, generating independent outputs $x_{i}$, $i=1,2,\ldots m$.
In this case we maximize $I(X^{m},\Theta)$ over $p(\theta)$ with
the goal of transmitting $\theta$ to high accuracy. The natural example
here is not telecommunications, but instead comes from viewing a scientific
experiment as a channel from the parameters $\theta$ in a theory,
via some noisy measurements, to recorded data $x_{i}$. Lindley \cite{Lindley:1956bj}
argued that the channel capacity or mutual information may then be
viewed as the natural summary of how much knowledge we will gain.
The distribution $p(\theta)$ is then a Bayesian prior, and Bernardo
\cite{Berger:1988vs,Zhang:1994ui} argued that the optimal $\popt(\theta)$
provides a natural choice of uninformative prior. This situation is
usually studied in the limit $m\to\infty$, where they named this
a reference prior. Unlike the case $m=1$ above, in this limit the
prior is typically continuous, and in fact usually agrees with the
better-known Jeffreys prior \cite{Clarke:1994gw}.

\begin{figure}
\centering \includegraphics[width=11cm]{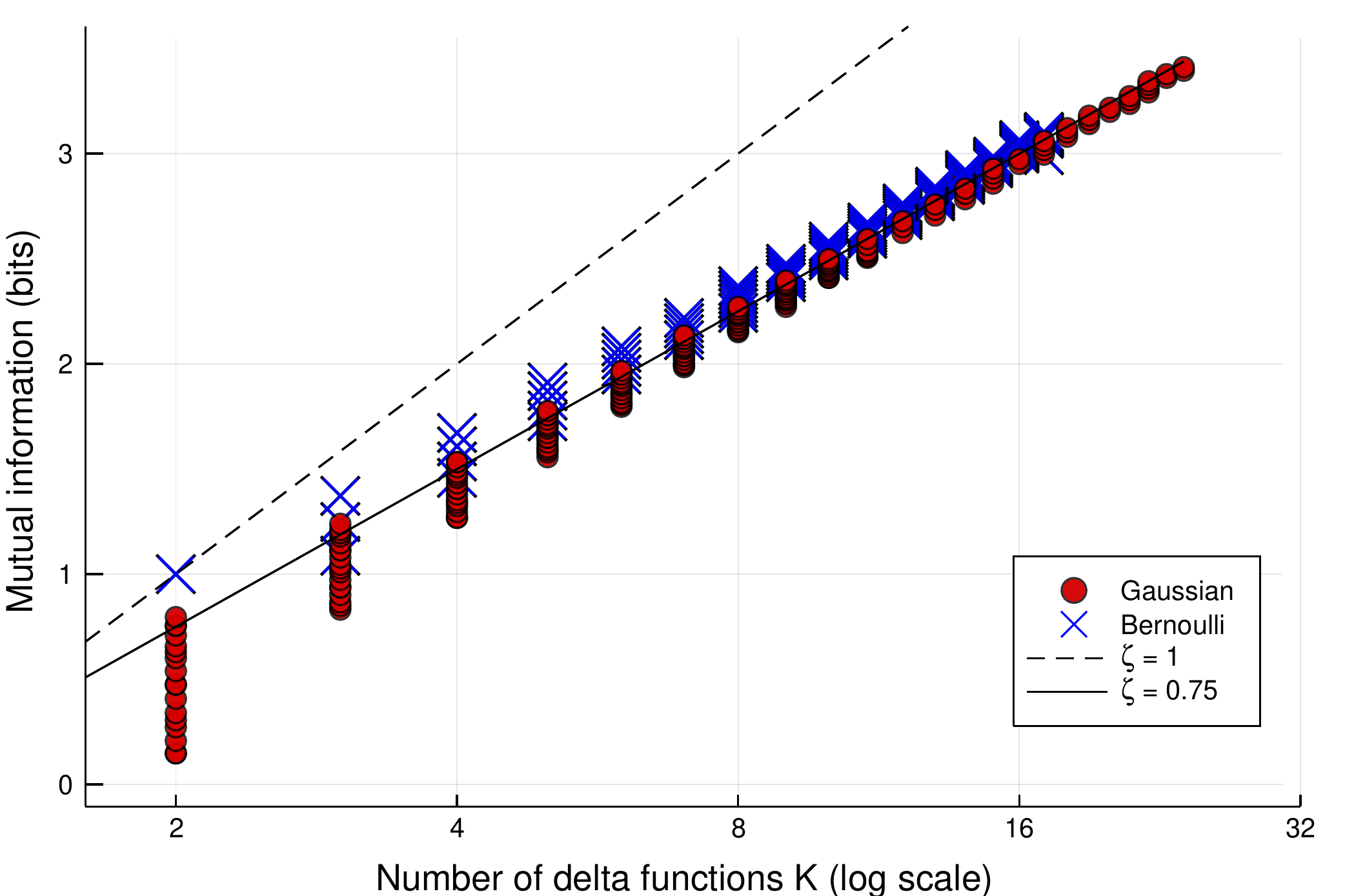}

\caption{Scaling law for the number of delta functions as $L$ is increased,
plotting $I(X;\Theta)/\log2$ against $K$. The lines drawn are $I(X;\Theta)=\zeta\:\log K$
for $\zeta=1$ (an absolute bound) and $\zeta=3/4$ (our scaling law).
The blue cross data points are for the Bernoulli model discussed in
section \ref{sec:3 Generalisations} below.\label{fig:Scaling-MI-K}}
\end{figure}

The result we report here is a novel scaling law, describing this
approach to a continuum. If $K$ is the optimal number of discrete
states, the form of our law plotted in Figure \ref{fig:Scaling-MI-K}
is this:
\begin{equation}
I(X;\Theta)\sim\zeta\log K\quad\text{ when }K\to\infty,\qquad\zeta=3/4.\label{eq:scaling-law-slope}
\end{equation}
Slope $\zeta=1$ on this figure represents the absolute bound $I(X;\Theta)\leq\log K$
on the mutual information, which simply encodes the fact that difference
between certainty and complete ignorance among $K=2^{n}$ outcomes
is exactly $n$ bits of information.

While the motivations above come from various fields, our derivation
of this law \eqref{eq:scaling-law-slope} is very much in the tradition
of physics. In Section \ref{sec:Derivation} we study a field theory
for the local number density of delta functions, $\rho(\theta)$.
The maximization gives us an equation of motion for this density,
and solving this, we find that the average $\rho_{0}=\int\rho(\theta)\:d\theta/L$
behaves as 
\begin{equation}
\rho_{0}=\frac{K}{L}\sim L^{-1+1/\zeta}=L^{1/3}\quad\text{ when }L\to\infty,\qquad\zeta=3/4.\label{eq:scaling-law-rho}
\end{equation}
Here $L$ is a proper length $L=\int\sqrt{g_{\theta\theta}\:d\theta^{2}}$,
measured with respect to the natural measure on $\Theta$ induced
by $p(x^{m}|\theta)$, namely the Fisher information metric \eqref{eq:ds2-FIM}.
At large $L$, this length is proportional to the number of distinguishable
outcomes, thus the information grows as $I(X;\Theta)\sim\log L$.
Then the scaling law \eqref{eq:scaling-law-slope} above reads $K^{\zeta}\sim L$,
equivalent to \eqref{eq:scaling-law-rho}.

We derive this law assuming Gaussian noise, but we believe that it
is quite general, because this is (as usual) a good approximation
in the limit being taken. Section \ref{sec:3 Generalisations} looks
explicitly at another one-dimensional model which displays the same
scaling (also plotted in Figure \ref{fig:Scaling-MI-K}) and then
at the generalization to $D$ dimensions. Finally appendix \ref{sec:Chang-and-Davisson}
looks at a paper from 30 years ago which could have discovered this
scaling law. But we begin by motivating in more detail why we are
interested in this problem, and this limit.

\section{Prior Work\label{sec:Motivation}}

How much does observing data $x$ inform us about the parameters $\theta$
of a model? Lindley \cite{Lindley:1956bj} argued that this question
could be formalized by considering the mutual information between
the parameters and the expected data, $I(X;\Theta)$. In this framework,
before data is seen we have some prior on parameter space $p(\theta)$,
and a resulting expectation for the probability of data $x$ given
by $p(x)=\int d\theta\:p(\theta)\:p(x|\theta)$. After seeing particular
data $x$, and updating $p(\theta)$ to $p(\theta|x)$ using Bayes'
rule, the entropy in parameter space will be reduced from $S(\Theta)$
to $S(\Theta|x)$. Thus on average the final entropy is $S(\Theta\vert X)=S(\Theta)-I(X;\Theta)$,
and we have learned information $I$. Bernardo and others \cite{Bernardo:1979uq}
argued that in the absence of any other knowledge, the prior $p(\theta)$
should be chosen to maximize $I$, so as to learn as much as possible
from the results of an experiment. The statistics community has mostly
focused on the limit where data is plentiful --- where each experiment
is repeated $m$ times, and $m$ goes to infinity. In this limit,
the prior which maximizes $I(X^{m};\Theta)$ for the aggregate data
$(x_{1},x_{2},\ldots,x_{m})$ is known as a reference prior \cite{Bernardo:1979uq}.
It usually approaches Jeffreys prior \cite{Clarke:1994gw}, which
can also be derived from invariance and geometric considerations,
described below.

In a recent paper \cite{Mattingly:2017uao} we argued that the finite
data case ($m\neq\infty$) contains surprises which naturally encode
a preference for model simplicity. This places model selection and
prior selection into the same framework. With finite data, it was
long known\footnote{There appear to be several independent discoveries of this fact in
the engineering literature \cite{Farber:1967us,Smith:1971kt,Fix:1978vk},
cited by different groups of later papers. Discreteness was also known
in related minimax problems \cite{Ghosh:1964ga,Casella:1981ex,Feldman:1991ba}.
In \cite{Mattingly:2017uao} we reviewed an argument for discreteness
from analyticity, and also demonstrated that the algorithm of \cite{Arimoto:1972jz,Blahut:1972ed},
known to be convex, converges to discrete points.} that the optimal prior $\popt(\theta)$ is almost always discrete,
composed of a finite number of delta functions: 
\begin{equation}
\popt(\theta)=\sum_{a=1}^{K}\lambda_{a}\:\delta(\theta-\theta_{a}).\label{eq:prior-sum-delta}
\end{equation}
The delta functions become more closely spaced as the number of repetitions
$m$ increases, with their density approaching the smooth Jeffreys
prior as $m\rightarrow\infty$, the limit of plentiful data. However,
in the data-starved limit, instead this prior has only small number
of delta functions, placed as far apart as possible, often at edges
of the allowed parameter space. It is the combination of these two
limiting behaviors which makes these priors useful for model selection.
The typical situation in science is that we have many parameters,
of which a few relevant combinations are in the data-rich regime,
while many more are in the data-starved regime \cite{Gutenkunst:2007gl,Machta:2013ga}.
If we are able to apply the methods of renormalization, then these
unmeasurable parameters are precisely the irrelevant directions. We
argued that, in general, $\popt(\theta)$ determines the appropriate
model class by placing weight on edges of parameter space along irrelevant
directions, but in the interior along relevant directions. Thus it
selects a sub-manifold of $\Theta$ describing an appropriate effective
theory, ignoring the irrelevant directions.

The appropriate notion of distance on the parameter manifold $\Theta$
should describe how distinguishable the data resulting from different
parameter values will be. This is given by the Fisher information
metric, 
\begin{equation}
ds^{2}=\sum_{\mu,\nu=1}^{D}g_{\mu\nu}d\theta^{\mu}d\theta^{\nu},\qquad g_{\mu\nu}(\theta)=\int dx\:p(x|\theta)\:\frac{\partial\log p(x|\theta)}{\partial\theta^{\mu}}\:\frac{\partial\log p(x|\theta)}{\partial\theta^{\nu}}\label{eq:ds2-FIM}
\end{equation}
which measures distances between points $\theta$ in units of standard
deviations of $p(x\vert\theta)$. Such distances are invariant to
changes of parameterization, and this invariance is an attractive
feature of Jeffreys prior, which is simply the associated volume form,
normalized to have total probability $1$: 
\[
\pjeff(\theta)=\frac{1}{Z}\sqrt{\det g_{\mu\nu}(\theta)},\qquad Z=\int d\theta\sqrt{\det g_{\mu\nu}(\theta)}.
\]
Notice, however, that normalization destroys the natural scale of
the metric. Repeating an experiment $m$ times changes the metric
$g_{\mu\nu}\to m\:g_{\mu\nu}$, encoding the fact that more data allows
us to better distinguish nearby parameter values. But this repetition
does not change $\pjeff(\theta)$. We argued in \cite{Mattingly:2017uao}
that this invariance is in fact an unattractive feature of Jeffreys
prior. The scale of the metric is what captures the important difference
between parameters we can measure well (length $L=\int\sqrt{g_{\theta\theta}\:d\theta^{2}}\gg1$
in the Fisher metric) and parameters which we cannot measure at all
(length $L\ll1$).

Instead, the optimal prior $\popt(\theta)$ has a different invariance,
towards the addition of extra irrelevant parameters. Adding extra
irrelevant parameters increases the dimensionality of the manifold,
and multiplies the volume form by the irrelevant volume. This extra
factor, while by definition smaller than $1$, can still vary by many
orders of magnitude between different points in parameter space (say
from $10^{-3}$ to $10^{-30}$). Jeffreys prior, and its implied distribution
$p(x)$, is strongly affected by this. It effectively assumes that
all parameter directions can be measured, because all are large in
the limit $m\to\infty$. But this is not true in most systems of interest
to science. By contrast $\popt(\theta)$ ignores irrelevant directions,
giving a distribution $p(x)$ almost unchanged by their addition,
or removal.

Bayesian priors have been a contentious subject from the beginning,
with arguments foreshadowing some of the later debates about wavefunctions.
Uninformative priors such as Jeffreys prior, which depend on the likelihood
function $p(x|\theta)$ describing a particular experiment, fit into
the so-called objective Bayesian viewpoint. This is often contrasted
with a subjective viewpoint, in which the prior represents our state
of knowledge, and we incorporate every possible new experimental result
by updating it: ``today's posterior is tomorrow's prior'' \cite{Lindley:1972tu}.
Under such a view the discrete $\popt(\theta)$, which is zero at
almost every $\theta$, would be extremely strange. Note however that
this subjective viewpoint is already incompatible with Jeffreys prior.
If we invent a different experiment to perform tomorrow, we ought
to go back and change our prior to the one appropriate for the joint
experiment, resulting in an update not described by Bayes' rule \cite{Lewis:2017uq}.
We differ only in that we regard (say) 10$^{6}$ repetitions of the
same experiment as also being a different experiment, since this is
equivalent to obtaining a much higher-resolution instrument.

But the concerns of this paper are different. We are interested in
the relevant directions in parameter space, as it is along such directions
that we are in the regime in which the scaling law \eqref{eq:scaling-law-slope}
holds. In the derivation of this, in Section \ref{sec:Derivation}
below, we study one such dimension in isolation. While we use the
notation of the above statistics problem, we stress the result applies
equally to problems from other domains, as discussed above.

\section{Derivation\label{sec:Derivation}}

This section studies a model in just one dimension, measuring $\theta\in[0,L]$
with Gaussian noise of known variance, thus 
\begin{equation}
p(x\vert\theta)=\tfrac{1}{\sigma\sqrt{2\pi}}\:e^{-(x-\theta)^{2}/2\sigma^{2}}.\label{eq:gaussian-model}
\end{equation}
We consider only one measurement, $m=1$, since more repetitions are
equivalent to having less noise. It is convenient to choose units
in which $\sigma=1$, so that $\theta$ measures proper distance:
$g_{\theta\theta}=1$. Thus $L$ is the length of parameter space,
in terms of the Fisher metric. Jeffreys prior is a constant $\pjeff(\theta)=1/L$.
The optimal prior has $K$ points of mass: 
\[
\popt(\theta)=\sum_{a=1}^{K}\lambda_{a}\:\delta(\theta-\theta_{a}),\qquad\theta_{1}=0,\quad\theta_{K}=L,\qquad\sum_{a=1}^{K}\lambda_{a}=1.
\]
The positions $\theta_{a}$ and weights $\lambda_{a}$ should be fixed
my maximizing the mutual information, This is symmetric $I(X;\Theta)=I(\Theta;X)$,
so can be written 
\begin{equation}
I(X;\Theta)=S(X)-S(X\vert\Theta)\label{eq:defn-MI}
\end{equation}
where the entropy and relative entropy are 
\begin{align*}
S(X) & =-\int dx\:p(x)\log p(x),\qquad p(x)=\int d\theta\:p(x\vert\theta)p(\theta)\\
S(X\vert\Theta) & =\int d\theta p(\theta)\left[-\int dx\:p(x\vert\theta)\log p(x\vert\theta)\right].
\end{align*}
For this Gaussian model, the relative entropy $S(X\vert\Theta)=\frac{1}{2}+\frac{1}{2}\log2\pi$
is independent of the prior, so it remains only to calculate the entropy
$S(X)$.

On an infinite line, the entropy would be maximized by a constant
$p(x)$, i.e. a prior with delta functions spaced infinitesimally
close together. But on a very short line, we observe that entropy
is maximized by placing substantial weight at each end, with a gap
before the next delta function. The idea of our calculation is that
the behavior on a long but finite line should interpolate between
these two regimes. We work out first the cost of a finite density
of delta functions, and then the local cost of a spatially varying
density, giving us an equation of motion for the optimum $\rho(x)$.
By solving this we learn how the density increases as we move away
from the boundary. The integral of this density then gives us $K$
with the desired scaling law.

\global\long\def\ripple{w}%
\global\long\def\move{\Delta}%
\global\long\def\height{h}%
\global\long\def\khat{\hat{k}}%
\global\long\def\pow{\eta}%
\global\long\def\SIGN{}%
 
\global\long\def\FACTOR{}%

Since the deviations from a constant $p(x)$ will be small, we write
\[
p(x)=\frac{1}{L}\left[1+\ripple(x)\right],\qquad\int dx\:\ripple(x)=0
\]
and then expand the entropy in powers of $\ripple(x)$:
\begin{align}
S(X) & =\log L-\frac{1}{2L}\int_{0}^{L}dx\:\ripple(x)^{2}+\bigo{\ripple^{4}}\nonumber \\
 & =\log L-\frac{1}{2}\sum_{k}\:\left|\ripple_{k}\right|^{2}+\ldots.\label{eq:S-int-lambda(k)}
\end{align}
Here our convention for Fourier transforms is that 
\[
\ripple_{k}=\int_{0}^{L}\frac{dx}{L}\:e^{-ikx}\ripple(x),\qquad k\in\frac{2\pi}{L}\mathbb{Z}.
\]

\subsection*{Constant spacing}

\begin{figure}
\centering\includegraphics[width=11cm]{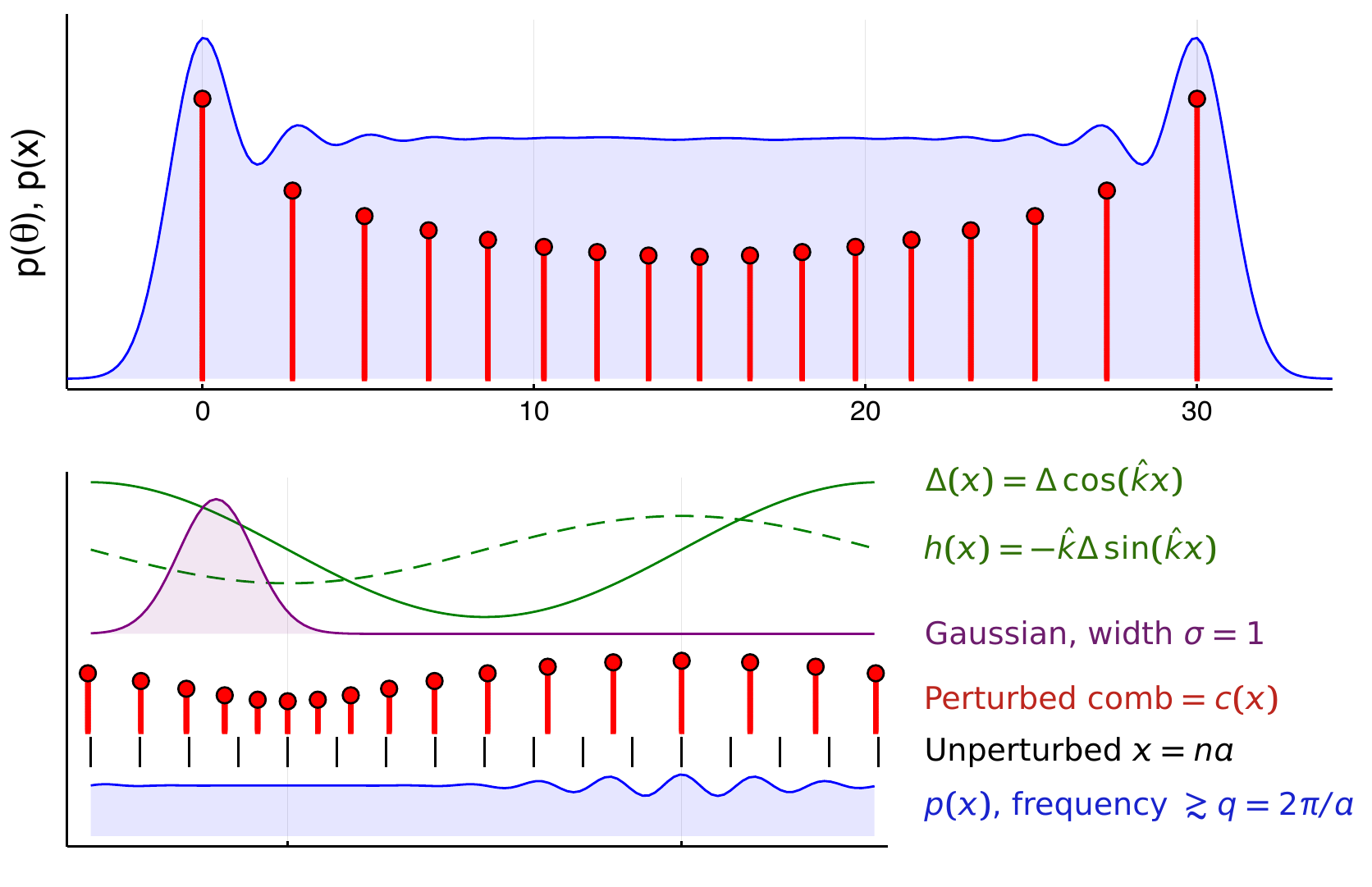}

\caption{Above, numerical solution $p_{\star}(\theta)$ for $L=30$, in which
we observe that as the spacing of the delta functions grows closer
together, their weights compensate to leave $p(x)$ almost constant,
with deviations $\protect\ripple(x)$ at a wavelength comparable to
the spacing. \medskip{}
\protect \\
 Below, a diagram to show the scales involved when perturbing the
positions of the delta functions in our derivation. These are arranged
from longest to shortest wavelength, see also \eqref{eq:all-scales}.
\label{fig:Example-and-Scales}}
\end{figure}

Consider first the effect of a prior which is a long string of delta
functions at constant spacing $a$, which we assume to be small compared
to the standard deviation $\sigma=1$, which in turn is much less
than the length $L$.\footnote{We summarise all the scales involved in \eqref{eq:all-scales}, see
also figure \ref{fig:Example-and-Scales}.} This leads to 
\begin{equation}
p(x)=\frac{a}{L}\sum_{n\in\mathbb{Z}}\:\frac{1}{\sqrt{2\pi}}e^{-(x-na)^{2}/2}.\label{eq:initial-p-of-x}
\end{equation}
Because this is a convolution of a Dirac comb with a Gaussian kernel,
its Fourier transform is simply a product of such pieces. Let us write
the transformation of the positions of the sources as follows: 
\[
c^{0}(x)=a\sum_{n\in\mathbb{Z}}\delta(x-na),\qquad\Rightarrow\qquad c_{k}^{0}=\frac{a}{L}\sum_{n\in\mathbb{Z}}e^{-ikna}=\sum_{m\in\mathbb{Z}}\delta_{k-m\frac{2\pi}{a}}.
\]
The zero-frequency part of $p_{k}$ is the constant term in $p(x)$,
with the rest contributing to $\ripple(x)$: 
\[
\ripple_{k}=\sum_{m\neq0}\:\delta_{k-m\frac{2\pi}{a}}\:e^{-k^{2}/2}=\begin{cases}
e^{-k^{2}/2} & k\in\frac{2\pi}{a}\mathbb{Z}\text{ and }k\neq0\\
0 & \text{else}.
\end{cases}
\]
The lowest-frequency terms at $k=\pm2\pi/a$ give the leading exponential
correction to the entropy: 
\begin{equation}
S(X)=\log L-e^{-q^{2}}+\bigo{e^{-2q^{2}}},\qquad q=\frac{2\pi}{a}.\label{eq:S-constant-density}
\end{equation}
As advertised, any nonzero spacing $a>0$ (i.e. frequency $q<\infty$)
reduces the entropy from its maximum.

\subsection*{Variable spacing}

Now consider perturbing the positions of the delta functions by a
slowly varying function $\move(x)$, and multiplying their weights
by $1+\height(x)$. We seek a formula for the entropy in terms of
$\move(x)$, while allowing $\height(x)$ to adjust so as to minimize
the disturbance. This cannot be done perfectly, as $\height(x)$ is
only sampled at spacing $a$, so only contributions at frequencies
lower than $q=2\pi/a$ will be screened. Thus we expect what survives
to appear with the same exponential factor as \eqref{eq:S-constant-density}.
In particular this ensures that at infinite density, no trace of $\move(x)$
remains. And that is necessary in order for the limit to agree with
Jeffreys prior, which is a constant.

Figure \ref{fig:Example-and-Scales} illustrates how the positions
and weights of $\popt(\theta)$ compensate to leave $p(x)$ almost
constant in the interior, in a numerical example. Below that it shows
how the functions $\move(x)$ and $\height(x)$ used here mimic this
effect.

The comb of delta functions $c^{0}(x)$ we had above is perturbed
to 
\[
c(x)=\left[1+\height(x)\right]\:a\sum_{n\in\mathbb{Z}}\delta\big(x-na-\move(na)\big)=\left[1+\height(x)\right]\:c^{\move}(x).
\]
The effect of $\height(x)$ is a convolution in frequency space: 
\[
c_{k}=c_{k}^{\move}+\sum_{k'}h_{k'}c_{k-k'}^{\move}.
\]
It will suffice to study $\move(x)=\move\cos(\khat x)$, i.e. frequencies
$\pm\khat$ only: $\move_{k}=\tfrac{1}{2}\move(\delta_{k-\khat}+\delta_{k+\khat})$.
The driving frequency is $\khat\ll q$. We can expand in the amplitude
$\move$ to write 
\begin{align*}
c_{k}^{\move} & =\frac{a}{L}\sum_{n}e^{-ik\big(na+\move(na)\big)}\\
 & =c_{k}^{0}-\frac{ik\move}{2}\left[c_{k-\khat}^{0}+c_{k+\khat}^{0}\right]-\frac{k^{2}\move^{2}}{8}\left[2c_{k}^{0}+c_{k-2\khat}^{0}+c_{k+2\khat}^{0}\right]+\bigo{\move^{3}}.
\end{align*}
The order $\move$ term has contributions at $k=\khat\ll q=2\pi/a$,
which can be screened in the full $c_{k}$ by setting $\height_{k}=+ik\move_{k}$
i.e. $h(x)=-\khat\move\sin(\khat x)$. What survives in $c_{k}$ then
are contributions at $k=0$, $k=\pm q$ and $k=\pm q\pm\khat$:\footnote{The contributions at and $k=\pm q\pm2\khat$ will only matter at order
$\move^{4}$ in $S(X)$.} 
\[
c_{k}=\delta_{k}+\sum_{\pm}\delta_{k\pm q}\Big(1-\frac{q^{2}\move^{2}}{4}\Big)+\sum_{\pm}[\delta_{k\mp q-\khat\SIGN}+\delta_{k\mp q+\khat\SIGN}]\Big(\pm\frac{i\,q\,\move}{2}\Big)+\bigo{\delta_{k\pm2q},\move^{2}}
\]
All but the zero-frequency term are part of $\ripple_{k}=(c_{k}-\delta_{k})e^{-k^{2}/2}$,
and enter \eqref{eq:S-int-lambda(k)} independently, giving this:
\begin{align}
\negthickspace\negthickspace S(X) & =\log L-e^{-q^{2}}\Big(1-\frac{q^{2}\move^{2}}{4}\Big)^{2}-\left[e^{-(q-\khat)^{2}}+e^{-(q+\khat)^{2}}\right]\Big(\frac{\move q}{2}\Big)^{2}+\bigo{\move^{4}}+\bigo{e^{-2q^{2}}}\nonumber \\
 & =\log L-e^{-q^{2}}\left[1+\move^{2}\left(q^{4}\khat^{2}+\frac{1}{3}q^{6}\khat^{4}+\frac{2}{45}q^{8}\khat^{6}+\ldots\right)\left[1+\bigo{1/q^{2}}\right]+\ldots\right]+\ldots\label{eq:S(X)-last}
\end{align}
As promised, the order $\move^{2}$ term comes with the same overall
exponential as in \eqref{eq:S-constant-density} above. Restoring
units briefly, the expansion in round brackets makes sense only if
$\khat\:q\:\sigma^{2}\ll1$.\footnote{In footnote \ref{fn:The-term-we-drop} we confirm that this indeed
holds.} In terms of length scales this means $\frac{2\pi/\khat}{\sigma}\gg\frac{\sigma}{a}$,
or writing all the assumptions made: 
\begin{equation}
\underset{\text{comb spacing}}{a=2\pi/q}\ll\underset{\text{kernel width}}{\sigma=1}\lll\underset{\text{perturbation wavelength}}{2\pi/\khat}\ll\underset{\text{box size}}{L}.\label{eq:all-scales}
\end{equation}

\subsection*{Entropy density}

We can think of this entropy \eqref{eq:S(X)-last} as arising from
some density: $S(X)=\int\frac{dx}{L}\mathcal{S}(x)=\mathcal{S}_{0}$.
Our claim is that this density takes the form 
\begin{equation}
\mathcal{S}(x)=\log L\SIGN-L\,e^{-(2\pi)^{2}\rho(x)^{2}}\left[1+\frac{128\pi^{6}}{3}\rho(x)^{4}\rho'(x)^{2}\right].\label{eq:entropy-density-rho'}
\end{equation}
The constant term is clearly fixed by \eqref{eq:S-constant-density}.
To connect the kinetic term to \eqref{eq:S(X)-last}, we need 
\[
\rho(x)=\frac{1}{a(1+\move'(x))}=\frac{1}{a}\left[1-\move'(x)+\move'(x)^{2}+\bigo{\move^{3}}\right]
\]
thus $\rho'(x)=-\frac{1}{a}\Delta''(x)+\ldots$ and 
\[
e^{-(2\pi)^{2}\rho(x)^{2}}=e^{-q^{2}}\left[1+2q^{2}\move'(x)+2q^{4}\move'(x)^{2}+\bigo{\move^{3}}\right]\left[1+\bigo{1/q^{2}}\right].
\]
Multiplying these pieces, the order $\move^{1}$ term of $\mathcal{S}(x)$
integrates to zero. We can write the order $\move^{2}$ term in terms
of Fourier coefficients (using \eqref{eq:S-int-lambda(k)}, and $\move'_{k}=ik\move_{k}$),
and we recover the leading terms in \eqref{eq:S(X)-last}. The next
term there $q^{8}\khat^{6}$ would arise from a term $\rho(x)^{6}\rho''(x)^{2}$
in the density, which we neglect.\footnote{The term $q^{8}\khat^{6}$ in $S(X)$ \eqref{eq:S(X)-last} corresponds
to a term $\rho^{6}(\rho'')^{2}$ in $\mathcal{S}(x)$ \eqref{eq:entropy-density-rho'}.
This gives a term in the equations of motion \eqref{eq:powers-of-x}
going like $x^{9\pow-4}$, which goes to zero as $x\to\infty$ with
$\pow=1/3$. Thus we are justified in dropping this. \label{fn:The-term-we-drop}}

\begin{figure}
\centering \includegraphics[width=11cm]{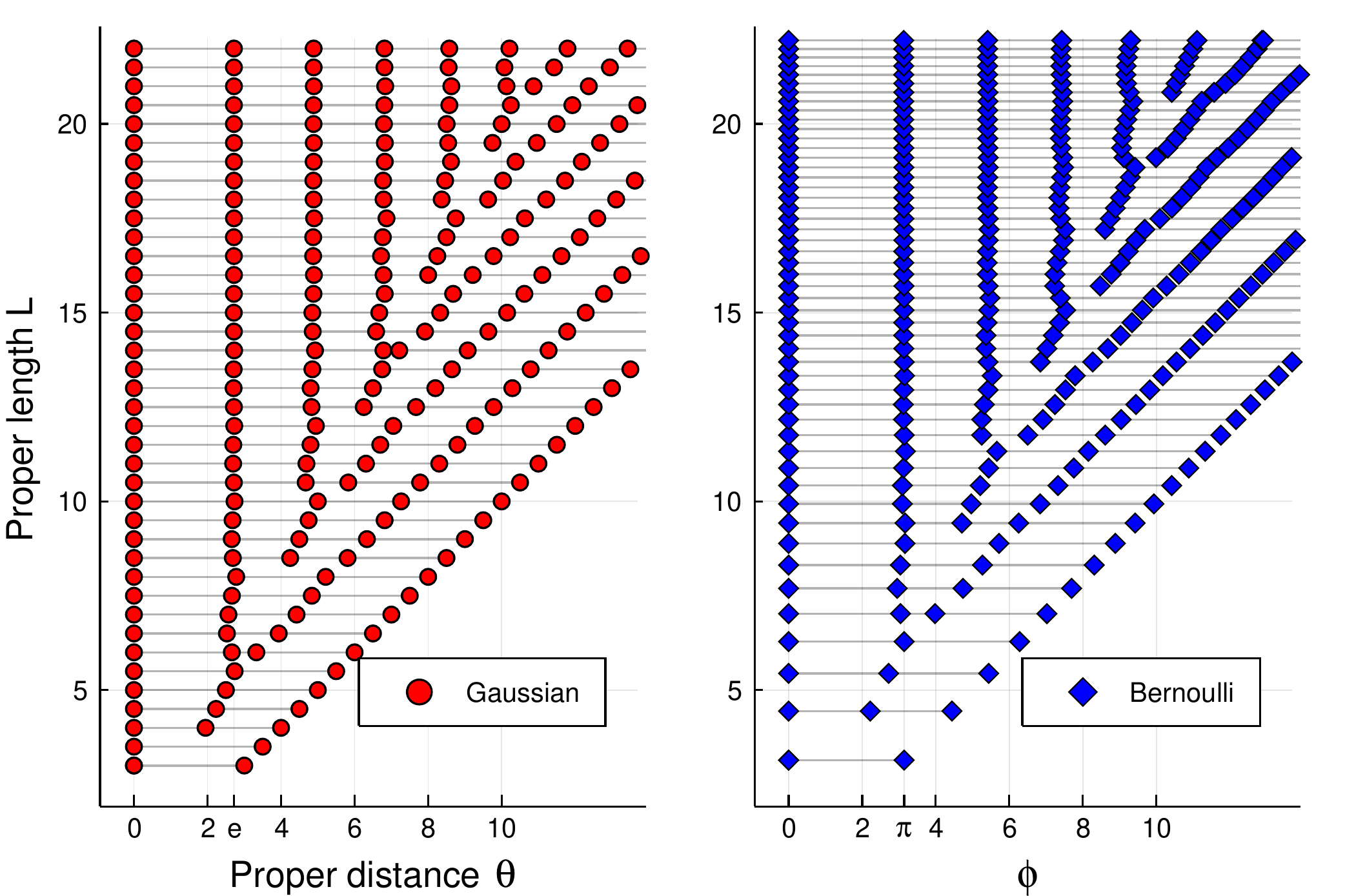}

\caption{Positions of delta functions in optimal priors $\popt(\theta)=\sum_{a=1}^{K}\lambda_{a}\:\delta(\theta-\theta_{a})$,
for various values of $L$. Each horizontal line represents one prior.
We observe that the second (and third...) delta functions occur at
fixed proper distance from the first, justifying the fixed boundary
condition on $\rho(x)$. \medskip{}
\protect \\
 On the right, we show similar data for the Bernoulli model of section
\ref{sec:3 Generalisations} below, in terms of proper distance $\phi$.
Here $L=\pi\sqrt{m}$ for $m=1,2,3,...50$. \label{fig:Atom-Positions}}
\end{figure}

The Euler-Lagrange equations from \eqref{eq:entropy-density-rho'}
read 
\[
0=\rho(x)^{4}\rho''(x)+2\rho(x)^{3}\rho'(x)^{2}-4\pi^{2}\rho(x)^{5}\rho'(x)^{2}+\frac{3}{32\pi^{4}}\rho(x).\FACTOR
\]
We are interested in the large-$x$ behavior of a solution with boundary
condition at $x=0$ of $\rho=1$. Or any constant density, but this
value is independent of $L$ because the only interaction is of scale
$\sigma=1$. This is also what we observe numerically, shown in figure
\ref{fig:Atom-Positions}. Making the ansatz $\rho(x)=1+x^{\pow}$
with $\pow>1$, these four terms scale as 
\begin{equation}
x^{5\pow-2},\quad x^{5\pow-2},\quad x^{7\pow-2},\quad x^{\pow},\qquad\text{all }\times\:e^{-x^{2\pow}},\;x\to\infty.\label{eq:powers-of-x}
\end{equation}
Clearly the first two terms are subleading to the third, and thus
the last two terms must cancel each other. We have $7\pow-2=\pow$
and thus $\pow=1/3$. Then the total number of delta functions in
length $L$ is 
\begin{equation}
K=\int_{0}^{L}dx\:\rho(x)\sim L^{4/3}\label{eq:K-integral-rho}
\end{equation}
establishing the result \eqref{eq:scaling-law-rho}.

\section{Extensions\label{sec:3 Generalisations}}

The other one-dimensional example studied in \cite{Mattingly:2017uao}
was Bernoulli problem, of determining the weighting of an unfair coin
given the number of heads seen after $m$ flips: 
\begin{equation}
p(x\vert\theta)=\frac{m!}{x!(m-x)!}\theta^{x}(1-\theta)^{m-x},\qquad\theta\in[0,1],\qquad x\in\{0,1,2,3,\ldots,m\}.\label{eq:bernoulli-model}
\end{equation}
The Fisher metric here is 
\[
g_{\theta\theta}=\frac{m}{\theta(1-\theta)}\qquad\Rightarrow\qquad L=\int_{0}^{1}\sqrt{g_{\theta\theta}\:d\theta^{2}}=\pi\sqrt{m}
\]
and we define the proper parameter $\phi$ by 
\[
ds^{2}=\frac{m\:d\theta^{2}}{\theta(1-\theta)}=d\phi^{2}\qquad\Leftarrow\qquad\theta=\sin^{2}\Big(\frac{\phi}{2\sqrt{m}}\Big),\quad\phi\in\left[0,\pi\sqrt{m}\right].
\]

The optimal prior found by maximizing the mutual information is again
discrete, and when $m\to\infty$ it also obeys the scaling law \eqref{eq:scaling-law-slope}
with the same slope $\zeta$. Numerical data showing this is also
plotted in figure \ref{fig:Scaling-MI-K} above. This scaling relies
on the behavior far from the ends of the interval, where this binomial
distribution can be approximated by a Gaussian: 
\begin{equation}
p(x\vert\theta)\approx\frac{1}{\sigma\sqrt{2\pi}}e^{-(x-m\theta)^{2}/2\sigma^{2}},\qquad m\to\infty,\;\theta\text{ finite},\qquad\sigma^{2}=m\:\theta(1-\theta).\label{eq:binomial-gaussian-approx}
\end{equation}
The agreement of these very different models suggests that the $\zeta=3/4$
power is in some sense universal, for nonsingular one-dimensional
models.

Near to the ends of the interval, we observe in figure \ref{fig:Atom-Positions}
that first few delta functions again settle down to fixed proper distances.
In this regime \eqref{eq:binomial-gaussian-approx} is not a good
approximation, and instead the binomial \eqref{eq:bernoulli-model}
approaches a Poisson distribution: 
\begin{equation}
p(x\vert\theta)\approx\frac{e^{-\mu}\mu^{-x}}{x!},\qquad m\to\infty,\quad\mu=m\theta\approx\frac{\phi^{2}}{2}\text{ finite}.\label{eq:binomial-poisson-approx}
\end{equation}
The first few positions and weights are as follows:\footnote{For the Gaussian model \eqref{eq:gaussian-model}, the corresponding
table reads: 
\[
\begin{aligned}\theta_{2} & \approx2.718 & \qquad &  & \lambda_{2}/\lambda_{1} & \approx0.672\\
\theta_{3} & \approx4.889 &  &  & \lambda_{3}/\lambda_{1} & \approx0.582\,.
\end{aligned}
\]
See also \cite{Amir:2016vi} and references therein.} 
\begin{align}
\qquad\phi_{2} & \approx3.13 & \lambda_{2}/\lambda_{1} & \approx0.63\qquad\nonumber \\
\phi_{3} & \approx5.41 & \lambda_{3}/\lambda_{1} & \approx0.54\label{eq:numerology}\\
\phi_{4} & \approx7.42 & \lambda_{4}/\lambda_{1} & \approx0.49\,.\nonumber 
\end{align}
This implies that the second delta function is at mean $\mu\approx2.47$,
skipping the first few integers $x$.

\subsection*{More dimensions}

Returning to the bulk scaling law, one obvious thing to wonder is
whether this extends to more dimensions. The trivial example is to
consider the same Gaussian model \eqref{eq:initial-p-of-x} in $D$-dimensional
cube: 
\begin{equation}
p(\vec{x}\vert\vec{\theta})\propto e^{-(\vec{x}-\vec{\sigma})^{2}/2},\qquad\vec{\theta}\in[0,L]^{D},\;\vec{x}\in\mathbb{R}^{D}.\label{eq:D-gauss}
\end{equation}
This simply factorizes into the same problem in each direction: \eqref{eq:defn-MI}
is the sum of $D$ identical mutual information terms. Thus the optimal
prior is simply 
\[
p_{\star}(\vec{\theta})=\prod_{\mu=1}^{D}p_{\star}(\theta_{\mu})=\sum_{a_{1}\ldots a_{D}=1}^{K}\lambda_{a_{1}}\cdots\lambda_{a_{D}}\:\delta(\theta-\theta_{a_{1}})\cdots\delta(\theta-\theta_{a_{D}})
\]
with the same coefficients as in \eqref{eq:prior-sum-delta} above.
The total number of delta functions is $K_{\mathrm{tot}}=K^{D}$ which
scales as 
\begin{equation}
K_{\mathrm{tot}}\sim L^{D/\zeta}=V^{1/\zeta},\qquad\zeta=3/4,\quad V\to\infty.\label{eq:multi-dim-scaling-law}
\end{equation}
We believe that this scaling law is also generic, provided the large-volume
limit is taken such that all directions expand together. If the scaling
arises from repeating an experiment $m$ times, then this will always
be true as all directions grow as $\sqrt{m}$.

To check this in a less trivial example, we consider now the bivariate
binomial problem studied by \cite{Polson:1990gm}. We have two unfair
coins whose weights we wish to determine, but we flip the second coin
only when the first coin comes up heads. After $m$ throws of the
first coin, the model is 
\begin{equation}
p(x,y\vert\theta,\phi)={m \choose x}\theta^{x}(1-\theta)^{m-x}{x \choose y}\phi^{y}(1-\phi)^{x-y}.\label{eq:bi-bi}
\end{equation}
with $\theta,\phi\in[0,1]$ and $0\leq y\leq x\leq m\in\mathbb{Z}$.
The Fisher information metric here is 
\[
ds^{2}=\frac{m}{\theta(1-\theta)}d\theta^{2}+\frac{m\,\theta}{\phi(1-\phi)}d\phi^{2}
\]
which implies 
\[
V=\int_{0}^{1}d\theta\:\int_{0}^{1}d\phi\:\sqrt{\det g_{\mu\nu}}=2\pi m
\]
and $p_{\mathrm{J}}(\theta,\phi)=\frac{1}{2\pi}\left[(1-\theta)\phi(1-\phi)\right]^{-1/2}$.
Topologically the parameter space is a triangle, since at $\theta=0$
the $\phi$ edge is of zero length. The other three sides are each
of length $\pi\sqrt{m}$, and so will all grow in proportion as $m\to\infty$.

\begin{figure}
\centering \includegraphics[width=11cm]{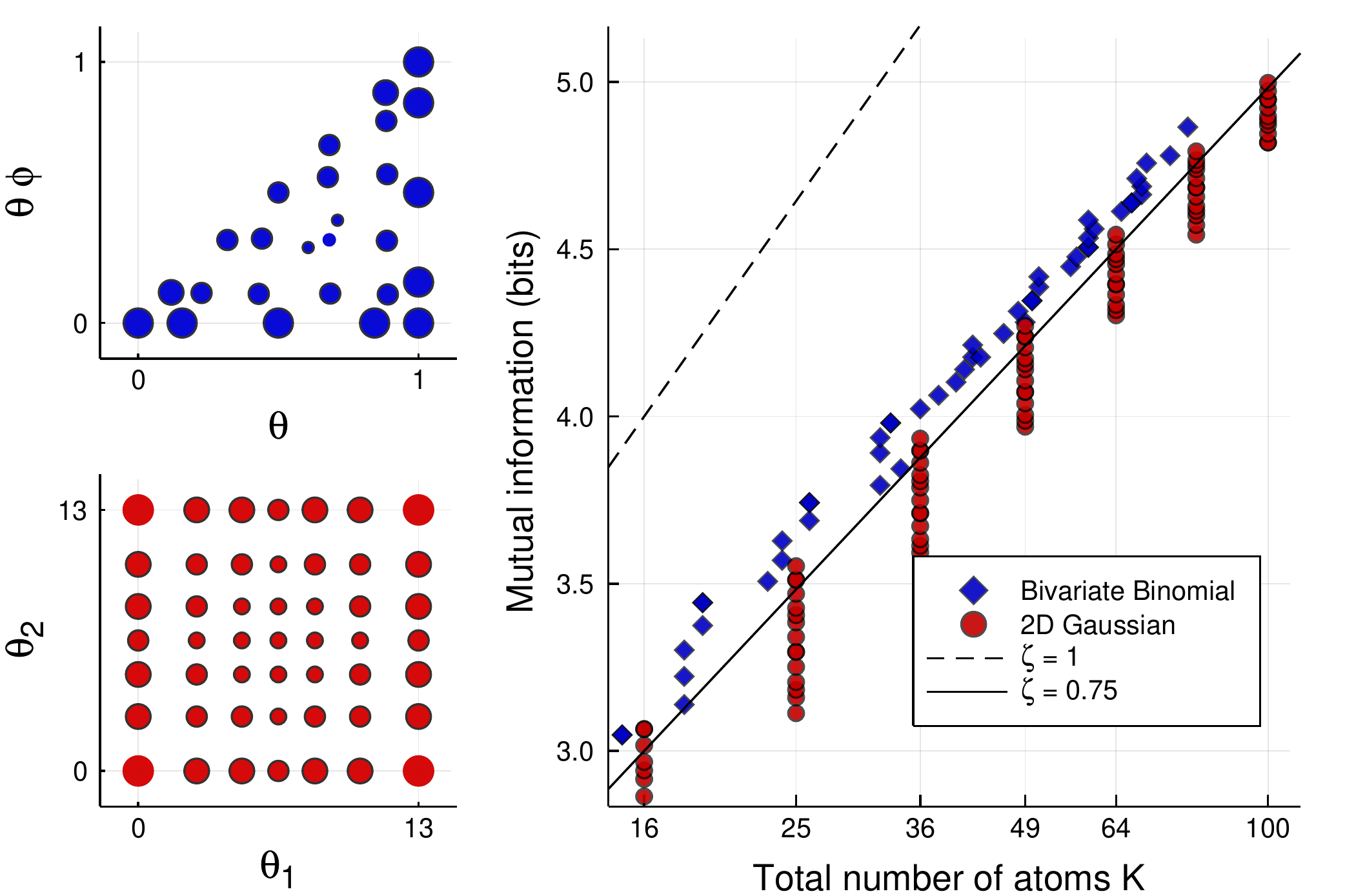}

\caption{Scaling law in two dimensions. On the right we plot $I(X;\Theta)/\log2$
against $\log K_{\mathrm{tot}}$ for the bivariate binomial model
\eqref{eq:bi-bi} and the $D=2$ Gaussian model \eqref{eq:D-gauss}.
On the left we show examples of the priors, for $m=20$ and $L_{1}=L_{2}=13$.
For the bivariate binomial the plot axes are $(\theta,\theta\phi)$
so as to respect the topology of the parameter space, but the figure
is not isometric to $\Theta$. \label{fig:Scaling-law-2D}}
\end{figure}

We can find the optimal priors for this numerically.\footnote{See the appendix of \cite{Mattingly:2017uao} for a discussion of
the algorithms used here, and \cite{Abbott:2017ri} for software.} In figure \ref{fig:Scaling-law-2D} we see that the mutual information
obeys the same law as \eqref{eq:scaling-law-slope} above: $I(X;\Theta)\sim\zeta\log K$
with $\zeta\approx0.75$. Since the Fisher volume is proportional
to the number of distinguishable states $I(X;\Theta)\sim\log V$,
this implies \eqref{eq:multi-dim-scaling-law}.

Finally, suppose that instead of a square (or an equilateral triangle),
a two-dimensional $\Theta$ has one direction much longer than the
other: 
\[
L_{1}=a_{1}\sqrt{m},\qquad L_{2}=a_{2}\sqrt{m},\qquad a_{1}\ll a_{2}.
\]
Then as we increase $m$ we will pass through three regimes, according
to how many of the lengths are long enough to be in the scaling regime:
\begin{align}
\qquad\qquad &  &  & L_{1},L_{2}\apprle1 &  & : & K & \text{ constant} &  & \qquad\qquad\nonumber \\
 &  &  & L_{1}\apprle1\ll L_{2} &  & : & K & \sim L^{1/\zeta}\propto m^{1/2\zeta}=m^{2/3}\label{eq:kinky-transitions}\\
 &  &  & 1\ll L_{1},L_{2} &  & : & K & \sim L^{2/\zeta}=V^{1/\zeta}\propto m^{1/\zeta}=m^{4/3}.\nonumber 
\end{align}
The last regime is the one we discussed above. When plotting $K$
against $\log m$ (or $\log L$), we expect to see a line with a series
of straight segments, and an increase in slope every time another
dimension becomes relevant.\footnote{These transitions are what LaMont and Wiggins call high-temperature
freeze-out \cite{LaMont:2017tk}.}

\section{Conclusion\label{sec:Conclusion}}

The fact that $\zeta<1$ is important for the qualitative behavior
of the priors studied in \cite{Mattingly:2017uao}. This is what ensures
that the number of delta functions $K\sim L^{1/\zeta}$ grows faster
than the Fisher length of parameter space $L$, ensuring that discreteness
washes out in the asymptotic limit $L\to\infty$. Parameters which
we can measure with good accuracy are in this limit. For such a parameter,
the posterior $p(\theta\vert x)$, which is also discrete, has substantial
weight on an increasing number of points, and in this sense approaches
a continuous description.

In section \ref{sec:3 Generalisations} we also studied some generalizations
beyond what we did in \cite{Mattingly:2017uao}. Very near to the
end of a long parameter, the discreteness does not wash out as $L\to\infty$,
and we wrote down its proper position for Gaussian and Poisson models.
And we observed that this scaling law holds in any number of dimensions,
if stated in terms of the mutual information \eqref{eq:scaling-law-slope}.
But stated in terms of the length $L$, it gives a slope which depends
on the number of large dimensions \eqref{eq:kinky-transitions}, and
hence has phase transitions as more parameters become relevant.

While our motivation here was finding optimal priors, our conclusions
apply to a much larger class of problems, including the maximization
of channel capacity over a continuous input distribution \cite{Arimoto:1972jz,Blahut:1972ed,Chang:1988bu,Lafferty:2001uj,Dauwels:2005vw},
which is formally equivalent to what we did above. This problem is
where mutual information was first discussed \cite{Shannon:1948wk},
and discreteness was first seen in this context \cite{Farber:1967us,Smith:1971kt,Fix:1978vk}.
This maximization is also equivalent to a minimax optimization problem
\cite{Haussler:1997fa}, and discreteness was known in other minimax
problems slightly earlier \cite{Ghosh:1964ga,Casella:1981ex,Feldman:1991ba};
see \cite{Berger:1988vs,Zhang:1994ui} for other work in statistics.
More recently discreteness has been employed in economics \cite{Sims:2006gu,Jung:2015uy},
and is seen in various systems optimized by evolution \cite{Balasubramanian:2009jx,Mayer:2015ce,Sharpee:2017fs}.
This scaling law should apply to all of these examples, when interpolating
between the coarse discreteness at small $L$ and the continuum $L\to\infty$.

\subsection*{Acknowledgements}

We thank Henry Mattingly and Mark Transtrum for collaboration on \cite{Mattingly:2017uao}.
This work was performed in part at Aspen Center for Physics, which
is supported by National Science Foundation grant PHY-1607611; M.C.A.'s
visit was supported by a grant from the Simons Foundation. M.C.A.
was supported by NCN grant 2012/06/A/ST2/00396, and a Wigner Fellowship.
B.B.M. was supported by a Lewis-Sigler Fellowship and by NSF PHY 0957573.

\appendix

\section{Chang and Davisson\label{sec:Chang-and-Davisson}}

\begin{figure}
\centering \includegraphics[width=11cm]{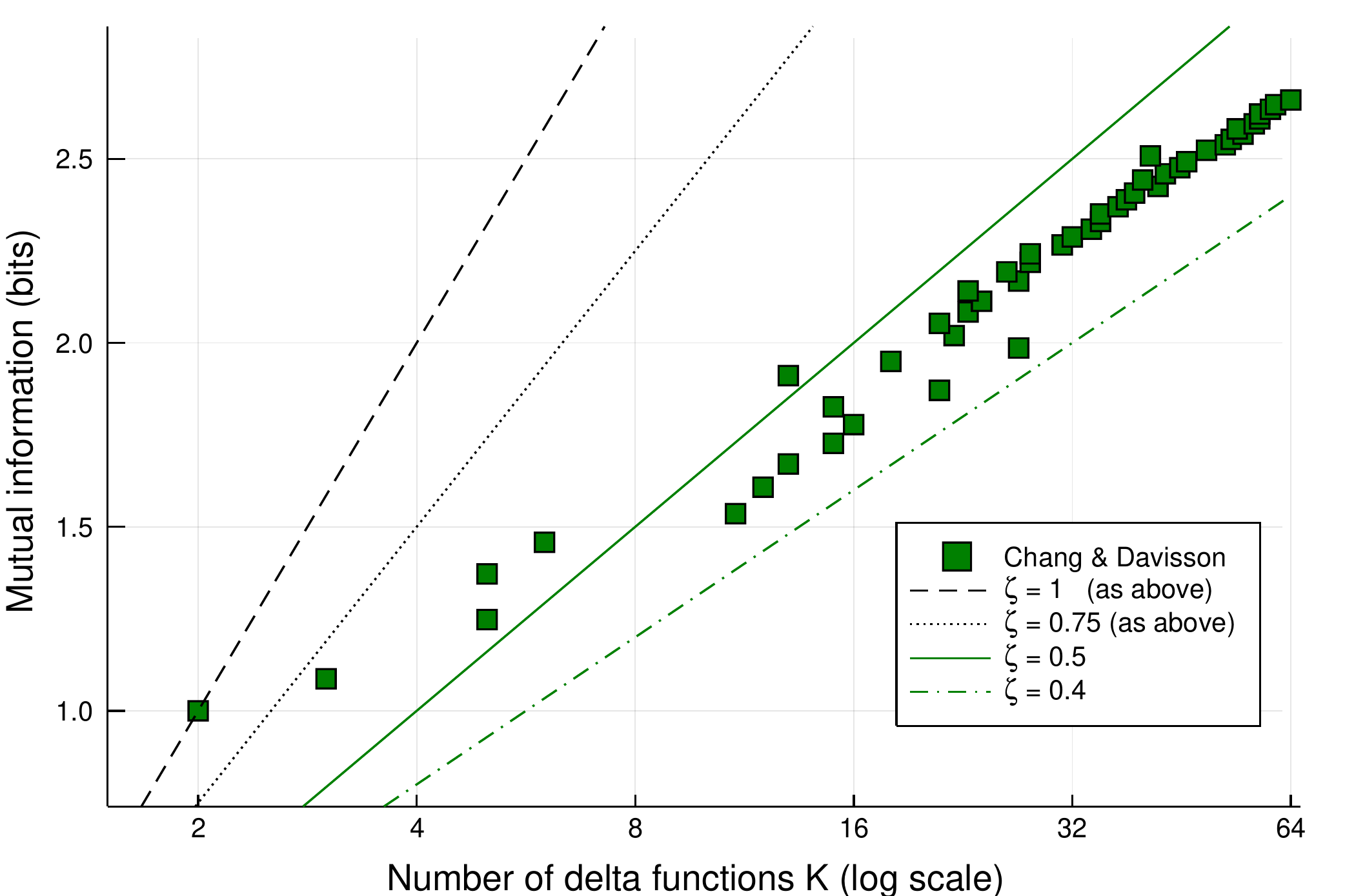}

\caption{Approximate scaling law \eqref{eq:CD-law} seen in the data from Table
1 of \cite{Chang:1988bu}. \label{fig:Chang-and-Davisson}}
\end{figure}

Chang and Davisson \cite{Chang:1988bu} came close to finding this
scaling law. In our notation, they maximized $I(X;\Theta)$ for various
$L$, and noted the number of delta functions $K'$ their algorithm
used to achieve good enough accuracy. But this is not quite the optimal
$K$: they tend to use too many delta functions, in a way which varies
with $L$. This results in a slightly different law: 
\begin{equation}
I(X;\Theta)\sim\zeta'\log K',\qquad\zeta'\apprle0.5\,.\label{eq:CD-law}
\end{equation}
Their paper does not mention this, nor attempt to plot this data as
in figure \ref{fig:Chang-and-Davisson}.

\bibliographystyle{my-JHEP-4}
\bibliography{for-MI-manual,for-MI-auto}

\end{document}